\documentclass[journal,twoside,web]{ieeecolor}
\usepackage{generic}
\usepackage{cite}
\usepackage{amsmath,amssymb,amsfonts}
\usepackage{algorithmic}
\usepackage{graphicx}
\usepackage{textcomp}

\usepackage{hyperref}       
\usepackage{url}            
\usepackage{booktabs}       
\usepackage{nicefrac}       
\usepackage{microtype}      
\usepackage{lipsum}
\usepackage{comment}
\usepackage{multirow}
\usepackage{breqn}
\usepackage{epstopdf}
\usepackage{cellspace}
\newcommand \mcE{\mathcal{E}}

\newcommand \mcG{\mathcal{G}}

\newcommand \mcV{\mathcal{V}}

\DeclareMathOperator*{\argmin}{argmin} 

\graphicspath{ {./images/} }

\def\BibTeX{{\rm B\kern-.05em{\sc i\kern-.025em b}\kern-.08em
    T\kern-.1667em\lower.7ex\hbox{E}\kern-.125emX}}
\begin{document}
\title{AI-based Identification of Most Critical Cyberattacks in Industrial Systems}
\author{Bruno P. Leao, Jagannadh Vempati, Siddharth Bhela, \IEEEmembership{Member, IEEE}, Tobias Ahlgrim, and Daniel Arnold, \IEEEmembership{Member, IEEE}
\thanks{Manuscript submitted 7 June 2023. This work was supported by the Cybersecurity, Energy Security, and Emergency Response (CESER), Risk Management Tools and Technologies
(RMT) Program of the U.S. Department of Energy via the Supervisory
Parameter Adjustment for Distribution Energy Storage (SPADES) Project and Mitigation via
Analytics for Grid-Inverter Cybersecurity (MAGIC) projects under
Contract DE-AC02-05CH11231. (Corresponding author: Bruno P. Leao.)}
\thanks{Bruno P. Leao, Jagannadh Vempati, Siddharth Bhela and Tobias Ahlgrim are with Siemens Technology, Princeton, NJ 08540 USA (e-mail: bruno.leao@siemens.com; jagannadh.vempati@siemens.com; siddharth.bhela@siemens.com; tobias.ahlgrim@siemens.com). }
\thanks{Daniel Arnold is with the
Lawrence Berkeley National Laboratory, Berkeley, CA 94720 USA (e-mail: dbarnold@
lbl.gov).}}

\maketitle

\begin{abstract}
Modern industrial systems face a growing threat from sophisticated cyberattacks that can cause significant operational disruptions. This work presents a novel methodology for identification of the most critical cyberattacks that may disrupt the operation of such a system. Application of the proposed framework can enable the design and development of advanced cybersecurity solutions for a wide range of industrial applications. Attacks are assessed taking into direct consideration how they impact the system operation as measured by a defined Key Performance Indicator (KPI). A simulation model (SM), of the industrial process is employed for calculation of the KPI based on operating conditions. Such SM is augmented with a layer of information describing the communication network topology, connected devices, and potential actions an adversary can take based on each device or network link. Each possible action is associated with an abstract measure of effort, which is interpreted as a cost. It is assumed that the adversary has a corresponding budget that constrains the selection of the sequence of actions defining the progression of the attack. A dynamical system comprising a set of states associated with the cyberattack (cyber-states) and transition logic for updating their values is also proposed. The resulting augmented simulation model (ASM) is then employed in an artificial intelligence-based sequential decision-making optimization to yield the most critical cyberattack scenarios as measured by their impact on the defined KPI. The methodology is successfully tested based on an electrical power distribution system use case.
\end{abstract}


\section{Introduction}
The design and implementation of advanced cybersecurity solutions is a very challenging task due to a number of reasons. The variety of possible cyberattack scenarios to consider can be enormous and they change over time as new vulnerabilities are identified and new hardware and software are integrated into systems. The availability of historical data associated with cyberattacks is also very limited, as they are relatively rare events and corresponding data, when collected, is usually confidential in nature. Cybersecurity of industrial systems poses even more challenges as each such system has in general very unique characteristics, for instance, in terms of hardware, software or operating conditions. They are also many times subject to increased exposure to vulnerabilities compared to information technology (IT) systems due to factors such as adoption of legacy communication protocols and challenges related to patching due to the associated down time. The need for advanced cybersecurity solutions for industrial systems is increasing faster than ever as Industry 4.0 trends materialize, especially the Industrial Internet of Things (IIoT) which expands the attack surface \cite{corallo2022}.

This work presents a novel solution that enables proper design and development of advanced cybersecurity solutions for industrial systems, especially focusing on cases where the goal of the adversary is the disruption of system operation. The proposed solution aims at identifying the most critical attack paths an adversary can take to disrupt specific key performance indicators (KPIs) of the industrial process. This information can then be employed to generate key test points for design and validation of advanced cybersecurity solutions.

Existing cybersecurity practices and tools have in general been adapted from their origins in IT to work on Operational Technology (OT) applications. They focus on the communication network perspective and take the system operation into account only indirectly, for instance, by defining abstractions such as critical assets which must be protected \cite{LeBlanc2017}. 

\begin{figure}[h]
    \centering
    \includegraphics[width=0.35\textwidth]{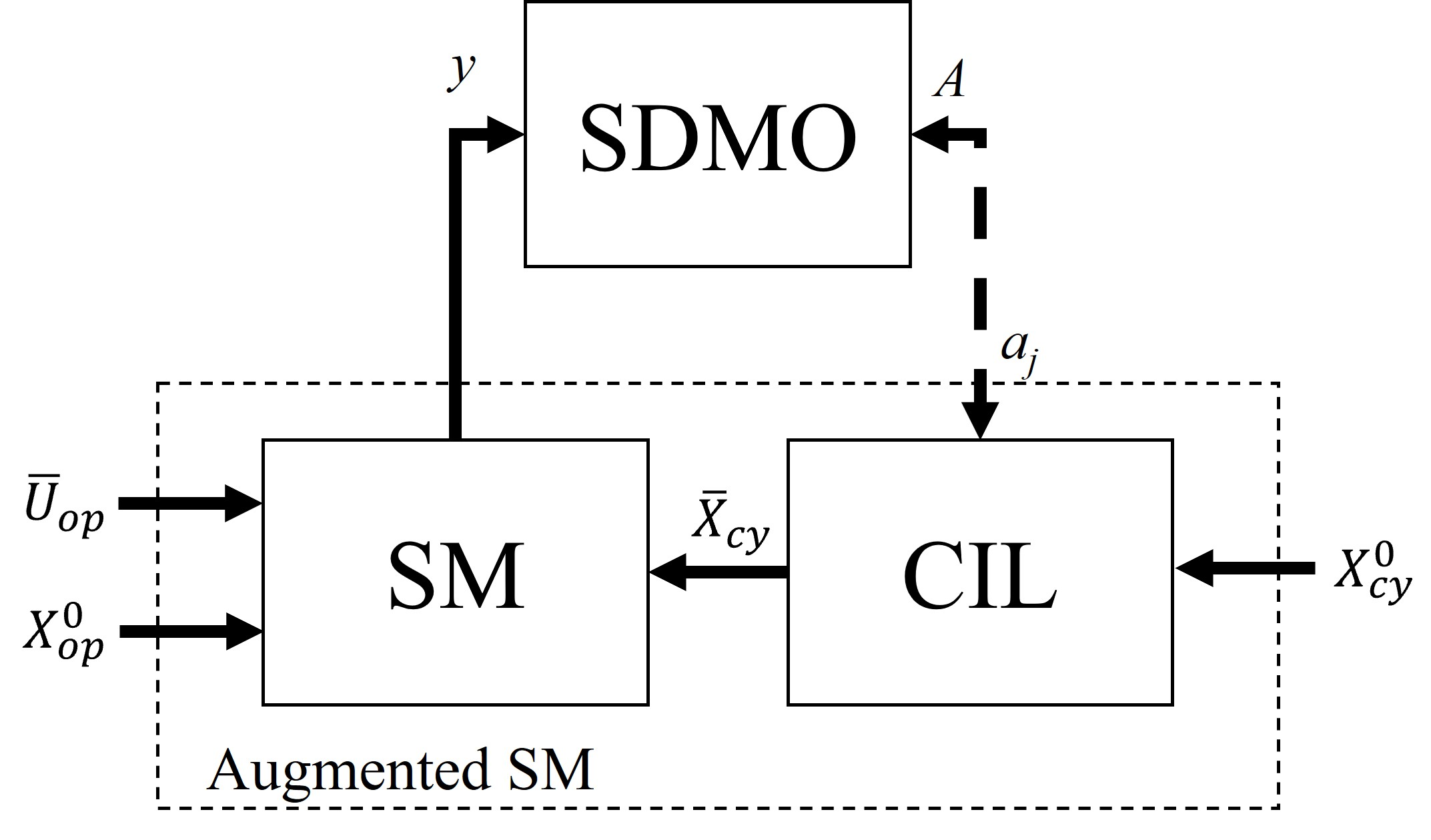}
    \caption{High-level structure of the proposed solution with three main components: simulation model of the industrial system (SM), cybersecurity information layer (CIL) and sequential decision-making optimization (SDMO). SM and CIL combined form the augmented SM (ASM). Solid arrows indicate information exchanged once for each evaluated attack scenario while the dashed arrow indicates information exchanged multiple times for each evaluated attack scenario. Definition of the symbols presented in the figure are described in section \ref{sec:augmented_dt}.}
    \label{fig:solution_diagram}
\end{figure}

The solution described here integrates the system operation explicitly into the cybersecurity assessment, providing a valuable analysis of possible threats in terms of their impact in such operation. This is achieved by means of a simulation model (SM) of the system which is employed to translate any operating conditions, including cyberattack situations, into their corresponding operational KPIs. Such system SM is augmented by a layer of information about the computer network and its cybersecurity characteristics, hereafter referred to as cybersecurity information layer (CIL). Through this augmentation, the relevant cyber-related aspects associated with the computation devices and communication links can be properly taken into consideration in the analysis of the physical system without the burden, complexity and computational costs associated with an emulation of the computation devices and network. Finally, a sequential decision-making optimization (SDMO) framework based on the Monte Carlo Tree Search (MCTS) method is employed to interact with the SM and CIL in order to identify the attack paths which cause the greatest damage to system operation. The diagram in Fig. \ref{fig:solution_diagram} depicts the three main components of the solution and their interrelation. 

The expertise required to apply the proposed methodology is the same as would be needed for existing cybersecurity risk assessment procedures, e.g. threat and risk assessment (TRA) or gap assessments, from an IT perspective. OT expertise, which by definition should be available for a certain industrial process, must be included to integrate the corresponding pieces of information. 

By employing this methodology, any set of relevant operational KPIs of an industrial system can be directly taken into consideration for properly adjusting the system's cybersecurity posture or designing cyber defense solutions. The resulting attack paths are defined in a realistic setting that includes the dynamics and constraints of real world adversary actions.

\subsection{Previous Work}
The proposed methodology can be compared to previous works in literature related to cybersecurity risk assessment and quantification and their application to industrial systems. Several risk assessment methods documented in the literature incorporate probabilistic analysis using Markov chains and Bayesian networks \cite{review_yohanandhan2020cyber}. Zhang et al. \cite{attack_graph_zhang2015power} explore four attack scenarios in SCADA networks, utilizing Bayesian attack graph models to assess the probabilities of successful cyber attacks. In this work, the authors utilize mean-time-to-compromise (MTTC) model  to estimate the time intervals for successfully intruding cyber components in control networks. Simulation results indicate that the system's reliability decreases as the frequency and sophistication of successful cyber attacks on the SCADA components increase.

Several techniques in this field also leverage attack graphs \cite{attack_graph_8720257, attack_graph_zhang2015power, attack_graph_stefanov2015scada, cao2018assessing, 9322321}. This graphical representation models the potential paths an adversary could take to compromise a system or network. Attack graphs identify the relationships between different vulnerabilities and attack vectors that an adversary could potentially exploit to gain unauthorized access to a target system. They represent an effective means of defining potential attack paths and has been widely used in the security evaluation of various systems, including critical infrastructure such as the electric grid. The attack graph methodology allows for the identification and prioritization of critical assets and can aid in improving the overall security posture of the system. The graph can be constructed using automated tools such as MulVAL \cite{ou2005mulval, ou2006scalable, tayouri2023survey} that map network topology, system configuration, and known vulnerabilities.  

Patapanchala PS et al. \cite{patapanchala2016exploring} propose a novel way to identify and prioritize critical assets and help operators take steps to improve the overall security of the system. In their work, they explore metrics that can be used to monitor the cybersecurity posture of the cyber-physical system (CPS) and the physical impact of an attack while considering individual and coordinated actions that can cause cascading outages. The authors demonstrate their metrics using cyber-physical models for 9-bus and 39-bus electrical grids. 

Kriaa et al. \cite{kriaa2015survey} conducted a comprehensive survey of existing approaches to industrial facility design and risk assessment that consider safety and security. They focused on the convergence of safety and security concerns in the context of the migration towards digital control systems, which creates new security threats that can endanger the safety of industrial infrastructures, and conducted a comparative analysis of related approaches identified in the literature. However, there are yet to be established methodologies for evaluating the influence of cybersecurity risks on higher-level processes and their corresponding organizational impact.


In \cite{cao2018assessing} a new method is proposed for assessing the impact of an attack on a business process-support enterprise network and generating a numerical score for the impact. The approach involves constructing a connected graph that maps out the dependencies between vulnerabilities on hosts, the relationships between services and hosts, and the dependencies between tasks and services. The authors leverage the MulVAL tool to generate the graph by encoding the dependencies with Datalog \cite{datalog_ceri1989you}, which is a declarative logic programming language. Finally, they quantify the impact based on the Common Vulnerability Scoring System (CVSS)\footnote{https://www.first.org/cvss/} scores of the vulnerabilities by propagating the information through the interconnected graph. CVSS is a widely used framework for assessing the severity of security vulnerabilities.


Haque et al. \cite{9322321} propose a graphical modeling technique that integrates mission-centric impact assessment of cyber attacks with a focus on operational resiliency. Their proposed approach combines the logical attack graph and mission impact propagation graph to calculate the impact of cyber attacks on the operational mission. Considering budgetary restrictions, an optimization process is also provided to minimize this effect. The authors demonstrate their modeling techniques through a case study involving SCADA systems for cyber-physical power systems. They suggest that by evaluating and minimizing the impact of cyber attacks on the operation of cyber-physical systems (CPS), their proposed method can enhance cyber resiliency.

The work proposed by Semertzis et al. \cite{semertzis2022quantitative} closely aligns with our proposed research. This work introduces a quantitative risk assessment method for cyber-physical systems, employing probabilistic and deterministic techniques. The method utilizes attack graphs to assess the likelihood of attacks and a dynamic cyber-physical power system model to simulate the impact of cyber attacks on power system cascading failures. A domain-specific language is proposed to describe digital substation assets and model attack graphs. The method calculates combined risk metrics considering the likelihood and impact of cyber threat scenarios. Experiments are performed using the IEEE 39-bus system, indicating that targeted cyber attacks on specific substations can result in significant cascading failures or even a blackout. 

The risk assessment methods mentioned above are mostly qualitative in nature or focus on employing probabilistic models to analyze the impact on a CPS. They often rely on small-scale networks and primarily focus on computer network perspectives, indirectly considering system operation through abstractions like critical assets. In contrast, our work focuses on quantitative risk assessment, pinpointing the most critical cyberattacks that can potentially disrupt industrial system operations and quantifying the impact of such disruptions. These attacks are directly assessed by evaluating their impact based on system operational KPIs. Our solution stands out by integrating system operation and network communication resulting in a comprehensive analysis of potential threats.  It is also agnostic to the type of OT operation, being applicable to a wide range of industrial systems. Based on the critical attack paths derived from the proposed solution, organizations can adequately prioritize their defense strategy and develop effective cyber-defense capabilities.

\subsection{Major Contributions}

The major contributions of this work are:
\begin{itemize}
    \item A novel methodology that provides a means for assessing the most critical cyberattacks aimed at disrupting the operation of industrial system as measured by defined operational KPIs.
    \item The means for integrating industrial system SMs into the analysis and augmenting them with the information required for implementation of the methodology.
    \item Empirical testing of the methodology with a realistic use case focused on a power distribution grid.
\end{itemize}

The remainder of the paper is organized as follows: section \ref{sec:methodology} contains all aspects of the methodology including the definitions associated with the augmented SM (ASM), both in terms of the additional information required and how it is processed, as well as the optimization problem formulation and proposed solution; in section \ref{sec:experiments} the experiments and corresponding results are presented and discussed; section \ref{sec:conclusion} corresponds to conclusion and future work.

\section{Methodology} \label{sec:methodology}

\subsection{Augmented Simulation Model (ASM)} \label{sec:augmented_dt}
The starting point of the methodology is a SM of the industrial system under analysis. Equation \eqref{eq:y_def} describes the main operation performed by the SM. $\Omega^\sigma$ corresponds to the information defining a cyberattack scenario. $y$ represents the KPI value of interest quantifying how well the system operates during such scenario. $y$ is a scalar variable which can be produced through a combination, e.g. weighted sum, of multiple indicators. The higher the value of $y$ the better the system is operating, therefore the goal of an adversary who wants to disrupt the system operation is to minimize $y$. Section \ref{sec:optimization_problem} presents a definition of the optimization problem.

The SM comprises a computational simulation of the industrial system of interest, implementing function $g_{SM}(.)$ \eqref{eq:y_def} used for calculating the KPI value $y^\sigma$ which is the outcome of attack scenario $\sigma$. In order to take into consideration the relevant information about network communication and cybersecurity, the SM is augmented by the CIL, resulting in the augmented SM (ASM) as presented in Fig. \ref{fig:solution_diagram}. Information about the progress of a cyberattack is represented in the form of a state vector containing what is hereafter referred to as cyber-states. Cyberattack scenario information $\Omega^\sigma$ \eqref{eq:Omega_def} is a set that includes two sequences and initial conditions for the system simulation ($X^0_{op}$) and cyber-states ($X^0_{cy}$), corresponding to all the information required for simulating the system of interest considering cyberattack scenario $\sigma$. The two sequences correspond to:
\begin{itemize}
    \item $\overline{U}^\sigma_{op}$: a sequence of inputs defining the normal operating conditions of the system \eqref{eq:u_bar}. 
    \item $\overline{X}^\sigma_{cy}$: a sequence of cyber-states defining the progression of an attack \eqref{eq:x_bar}. Only a subset of the cyber-states is relevant for the SM, consisting of those related to the adversary actions which have an impact in the industrial system of interest. They correspond to actions of category \textit{impact} as explained in section \ref{sec:actioncategories}. 
\end{itemize}

Not only the sequence but also the simulation times corresponding to changes in operating conditions and execution of adversary actions may affect the KPI results. Those times are considered in the proposed framework but they are not included in the notation for simplicity.
 
The SM is employed as a black-box in the optimization process which makes the solution agnostic to the application domain or SM characteristics. 
\begin{equation} \label{eq:Omega_def}
\Omega^\sigma = \{\overline{U}^\sigma_{op}, \overline{X}^\sigma_{cy}, X^0_{op}, X^0_{cy}\}
\end{equation}
\begin{equation} \label{eq:u_bar}
\overline{U}^\sigma_{op} = [U^\sigma_{op,1}, U^\sigma_{op,2}, ..., U^\sigma_{op,z^\sigma_{op}}]
\end{equation}
\begin{equation} \label{eq:x_bar}
\overline{X}^\sigma_{cy} = [X^\sigma_{cy,1}, X^\sigma_{cy,2}, ..., X^\sigma_{cy,z^\sigma_{cy}}]
\end{equation}
\begin{equation} \label{eq:y_def}
y^\sigma = g_{SM}(\Omega^\sigma)
\end{equation}

The CIL comprises two parts: 
\begin{enumerate}
    \item a collection of information about each device and communication network link in the system, hereafter referred to as cybersecurity information (CI). This is specific to the industrial application under consideration.
    \item an engine for sequentially updating cyber-states given adversary actions ($a_j$), providing the sequence of cyber-states defining an attack scenario ($\overline{X}_{cy}$), and identifying all possible actions the adversary can take ($A$) given a certain cyber-state. This engine is hereafter referred to as cybersecurity information engine (CIE) and it is application agnostic.
\end{enumerate}

The subsections below present relevant concepts concerning adversary action categories, costs and virtual links, followed by a detailed explanation about the CI and the CIE.

\subsubsection{Categories of Adversary Actions} \label{sec:actioncategories}

In the proposed methodology, possible actions an adversary can take during an attack are grouped into three categories: \textit{access}, \textit{exploit} and \textit{impact} which are explained in detail below. The definition of each of them is in direct alignment with MITRE ATT\&CK \cite{mitre_attack} tactics, techniques and procedures (TTPs). The choice of grouping multiple TTPs into three categories was made to keep the framework as simple as possible, facilitating its understanding and application.

\paragraph{Access} \label{par:access}
The \textit{access} category corresponds to the efforts required for accessing the target system either externally - from the internet or creating a physical security breach - or internally - by means of lateral movement. Examples of TTPs which map to this category include:
\begin{itemize}
    \item Network Scanning: An adversary may use network scanning to identify exposed systems or services that can be targeted for exploitation.
    \item Phishing: An adversary may use phishing to trick users into providing credentials or clicking on malicious links, allowing the adversary to gain access to the network.
\end{itemize}

\paragraph{Exploit}\label{par:Exploit}
Actions in this category correspond to technical vulnerabilities that can be exploited by the adversary. Such actions may be associated, for instance, with software vulnerabilities and misconfigurations. TTPs associated with this category include the following:
\begin{itemize}
    \item Remote Command Execution: This technique involves an adversary obtaining means for remotely executing commands on a targeted industrial control system. By exploiting vulnerabilities or leveraging authorized remote access, the adversary gains control over the system and can execute malicious commands to manipulate or disrupt its operation.
    \item Rogue Master: Adversaries have the capability to establish a rogue master that takes advantage of control server functionalities to communicate with outstations. This rogue master can be utilized to send control messages that appear legitimate to other devices, enabling actions that cause impacts on the industrial processes.
\end{itemize}

\paragraph{Impact}
This category comprises adversary actions which directly impact the operation of the target industrial system. Concerning the methodology, it comprises actions that may affect the SM simulation and therefore may impact the KPI of interest ($y$). Some examples of TTPs which can be mapped to this category include:
\begin{itemize}
    \item Impact to Availability: An adversary may launch a denial of service attack or destroy critical data, disrupting services and causing downtime.

    \item Impact to Integrity: An adversary may modify data or inject malicious code, compromising the integrity of the system or network.
\end{itemize}

\subsubsection{Action Costs and Attack Budget} 
\label{sec:actioncosts}

As part of the proposed methodology, each action is associated with an abstract measure of effort\footnote{Although the costs are presented here as a measure of effort, they could be easily adapted to represent any other quantity that would be accumulated - e.g. summed - for each action taken by the adversary such that this accumulated value can be used to define the end of the attack - e.g. by comparison with a threshold. Examples of additional possibilities for cost in this case include time or some "detectability" measure. Plans for future work include the exploration of related alternatives.}, referred hereafter to as action costs or simply costs. The costs associated with each action have an important role in the methodology as they define the constraints which limit the attack scenarios that can be produced. 

A predefined budget is defined as a parameter in the attack optimization so that the summation of the costs of all actions taken in a certain attack scenario cannot exceed such budget. As the structure of the proposed framework is modular, information defined at the component or network link level, including costs, can to a large extent be reused, facilitating the application of the methodology for various use cases.

Previous works in the literature consider a budget definition for cyberattack and defense, which is evidence of the usefulness of this kind of approach. However, such existing definitions are considered insufficient for the purpose of this work. For instance, in \cite{hasan2018vulnerability} each node in a power grid is assumed to be a substation and attack and defense budgets are simply defined in terms of a number of substations. In \cite{Hasan2020} a game-theoretical approach to attacking is described, with worst-case and random attacks utilized, but the attack and defense budgets are also defined in terms of a number of substations. The attack budget and corresponding action costs defined in this work extend these ideas to take into consideration the trade-offs and decision making associated with a large variety of possible adversary actions.

Below is a description of the subjective factors influencing the definition of costs for each action category followed by a discussion of objective means for quantification of the corresponding values as required for implementing the methodology.

\begin{itemize}
    \item \textit{Access}: Costs in this category are associated with factors such as network topology, access controls, and security posture. High costs in this case could correspond to devices or systems that are hard to access and are well protected. Low costs on the other hand could result, for instance, from devices or systems that are directly connected to the internet with little or no protection. As the costs for actions in this category may depend to a large extent on factors such as physical locations and cyber defenses, values defined for one application may not be as reusable for other applications as the ones from the other categories.
    \item \textit{Exploit}: The corresponding costs reflect the complexity associated with compromising the system to achieve the attack objectives. High costs associated with an action in exploit category could correspond to a need for specialized knowledge or even the need for a new zero-day exploit. Low costs on the other hand may be associated with known exploits that are easy to use. The costs defined in this category have a great potential of reuse as they refer to the internal vulnerabilities associated with devices and network links which should be similar for different applications.
    \item \textit{Impact}: The definition of costs in this category will in general be very dependent on the characteristics of the OT system and device, so the definition of costs will benefit from the support of domain experts. Low costs may be associated, for instance, with modification of configuration parameters that can be directly accessed by a user with proper privileges. High costs may be associated with actions requiring, for instance, changes in the device firmware to override built-in protections. Cost values defined in this category have a high potential of reuse as they refer to operations affecting the internals of a device or network link which should be similar for different applications.

\end{itemize}

\paragraph{Action Costs Quantification} 
\label{sec:costsquantification}

Proper quantification of the costs associated with attacker's actions is essential for realistic attack scenario modeling. The following are relevant aspects which must be taken into consideration when defining a method for such quantification:
\begin{itemize}
    \item The decision making defined as part of the methodology requires choosing among actions which can be in any of the three different categories defined above. Therefore, the quantification of action costs must be consistent across those categories.
    \item The absolute values of the costs are not relevant for the decision making. Only the relative values matter. As presented in section \ref{sec:optimization_problem}, costs along with the corresponding budget define constraints for an optimization problem. It can be noticed from \eqref{eq:constraint} that scaling the costs by any positive factor does not affect the constraint, as long as the budget is also scaled according to the same factor.
\end{itemize}

Directly defining the costs for each attack in an arbitrary scale and range may be an option when the list of all possible attacks is not too extensive and proper subject matter experts are available so that an analysis can be manually performed taking into account all of the attacks. This is especially feasible when simplification assumptions can be performed to create a reduced number of cost groups for each attack category. Considering a case where multiple different action possibilities exist in a certain attack category, the analysis of an expert may indicate, for instance, that all those actions can be grouped into two cost levels, corresponding to low and high effort.

Methods such as the Delphi Technique \cite{okoli2004delphi} can also be employed to structure the expert's input process, aiming to mitigate biases and promote consensus. In this case, the first step is to establish a quantitative cost scale. Next, each expert anonymously assigns cost values to each action. In subsequent rounds, the median of cost estimates are summarized and presented anonymously. The experts then have the opportunity to revise their estimates based on the collective feedback provided in the form of such summarized scores. Multiple rounds can be conducted until reaching a level of consensus which can be quantitatively defined as a function of the difference between the medians obtained from subsequent rounds, i.e. if this difference is below a defined threshold it is considered that consensus has been achieved.

As direct definition of the costs may not be effective or feasible in many situations, an alternative approach is proposed here which is based on the application of the Elo rating system \cite{glickman1999}. This is the main system employed in the last decades for rating chess players. It has also been employed for various other purposes, including a recent application in the comparison of large language models performance \cite{zheng2023judging}. Concerning the application of this method for definition of the action costs, the various adversary action possibilities are compared in terms of the associated relative effort and the resulting rating numbers are employed as the cost values. This approach is simple and compatible with the two relevant aspects presented above. It also has one characteristic that makes it scalable and flexible: rating updates are based on pairwise comparisons. This means that the expert performing the analysis does not need to take into consideration all the actions at the same time. The approach only requires the comparison of two actions at a time, with a simple outcome indicating that one of them requires more, similar or less effort than the other. Each pairwise comparison is performed independently of the others. This also enables sharing of the work among different experts if so desired. Another benefit of this approach is that the ratings can be reused for different applications, with a straightforward process for addition or removal of adversary actions to/from the list of possibilities that does not require additional SME effort for re-quantifying other costs. 

Equations \eqref{eq:elorating} and \eqref{eq:eloupdate} describe the process for updating the cost $\phi_A$ associated with action $A$ based on a comparison with action $B$ with corresponding cost $\phi_B$. $E_A$ is an estimate of the probability of action $A$ being considered to have a similar or higher cost when compared to action $B$. $G_A$ is the outcome of the comparison between the two actions, resulting in one of the values 1, 0.5 or 0 corresponding respectively to action $A$ being considered more, similarly or less costly than action $B$. Superscripts minus ($-$) and plus ($+$) correspond respectively to the costs before and after the update. $K_E$ and $K_u$ are gains which can be tuned to adjust the range and variability of the scores.

\begin{equation} \label{eq:elorating}
    E_A = \frac{1}{1+10^{(\phi_B-\phi_A)/K_E}}
\end{equation}
\begin{equation} \label{eq:eloupdate}
    \phi_A^+ = \phi_A^- + K_u(G_A - E_A)
\end{equation}

Differently from chess ratings, the time when the comparison is performed is not relevant for the action cost definition. Therefore, a bootstrap method can be employed as proposed in \cite{zheng2023judging} to produce values that do not depend on the order the comparisons are made. Grouping actions into cost levels, as suggested above for the direct definition of costs, can potentially also improve the efficiency in this case by reducing the number of pairwise comparisons required for achieving meaningful cost values.  

Although the methods described above provide systematic means for defining the cost values, they also present limitations. Ideally, those values should be parameterized to account for dependencies on other actions, e.g. execution of exploit $A$ may reduce the cost of executing exploit $B$, or system changes, e.g. modifications in cyber defenses. Well accepted cybersecurity rating systems such as the CVSS could also be leveraged for definition of the costs. Those aspects will be investigated as part of future work.

\subsubsection{Logical Links and Network Links} \label{sec:logiclinks}

\begin{figure} [h]
    \centering
    \includegraphics[width=0.48\textwidth]{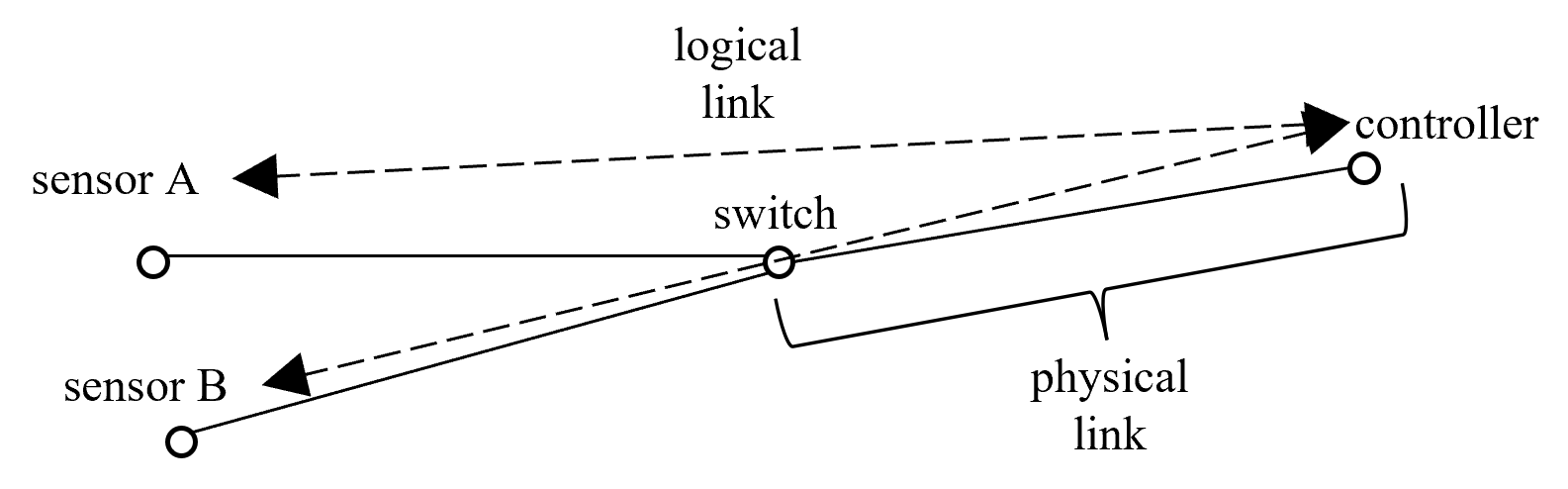}
    \caption{Diagram representing the interconnection among devices. Solid lines indicate network (physical) links and dashed lines correspond to logical links.}
    \label{fig:logiclink}
\end{figure}

A logical link is defined here as a virtual communication channel between two devices that establish a path for data or information exchange,  independently of the underlying network infrastructure.  Data or information in this case includes only what is relevant for the system operation. An example can be described based on an industrial communication network where sensors and controllers are connected by means of network switches. In this case, physical computer network links, referred to in this work as network links, exist between controllers and switches as well as between sensors and switches. However, the switches are not part of any logical link as they do not produce or consume the data relevant for the system. One network link may be associated with multiple logical links and a logical link may as well be associated with multiple network links. Fig. \ref{fig:logiclink} presents a more concrete example with two sensors, $A$ and $B$, connected to a controller by means of a network switch. Network links connect the sensors and the controller to the switch as indicated by the solid lines in the diagram. There is no network link connecting sensors directly to the controller. Both sensors produce information which is consumed by the controller, therefore there are two logical links connecting the sensors to the controller as indicated by the dashed arrows. The switch is not associated with any logical link. In this example, the network link connecting the controller to the switch is associated with two logical links and each logical link is associated with two network links.

The distinction between network link and logical link is relevant as different aspects of an attack may depend on one or the other. For instance, aspects related to actions of type \textit{access} may depend on the physical location of the network link, while the effects of actions of type \textit{impact} will depend on the related logical links.

\subsubsection{CI} \label{sec:ci}
The CI comprises a set of specifications for each device and for each network link composing the system. Equation \eqref{eq:ci_def} defines the CI as a pair of sets $S^d$ and $S^l$, corresponding respectively to specifications of devices and network links.

\begin{equation} \label{eq:ci_def}
CI = (S^d, S^l)
\end{equation}

Equations \eqref{eq:ci_dev} to \eqref{eq:action_param_dev} describe the specifications associated with devices as indicated by the $d$ superscript. In such equations, $s^{d_i}$ corresponds to the specifications for the $i^{th}$ device, $n^d$ being the total number of devices. Specifications for device $i$ correspond to its type $\tau^{d_i} \in T$ and $\alpha^{d_i}$ which is the set of all actions an adversary can take when interacting with it. There is a total of $m^{d_i}$ possible actions for an adversary to interact with $d_i$, although many of those actions may not be accessible under many circumstances as explained in section \ref{sec:cie}.

Each possible action $a^{d_i}_j$ is defined as a set that includes an action type $\psi^{d_i}_j \in \Psi(\tau^{d_i})$, a set of parameters $\theta^{d_i}_j$ which are a function of the action type, and the cost $\phi^{d_i}_j$ which is an abstract measure of the effort required for executing the corresponding action as explained in section \ref{sec:actioncosts}. 

The set of parameters $\theta^{d_i}_j$ includes a category $\gamma^{d_i}_j$ which can be one of \textit{access}, \textit{exploit} or \textit{impact}, as explained in section \ref{sec:actioncategories}. The definition of what additional parameters will constitute $\theta^{d_i}_j$ depend on the corresponding action category: $e^{d_i}_j$ is an indicator function which is 1 if the action of category \textit{access} corresponds to an entry point and 0 otherwise. $\zeta^{d_i}_j$ and $\xi^{d_i}_j$ are respectively a set of \textit{impact} attacks which become available and a set of other devices which become visible following execution of an action of category \textit{exploit}. Finally, $\pi^{d_i}_j$ is the mapping to a function which is part of the SM and defines the effect in simulation associated with an action of category \textit{impact}.

\begin{equation}\label{eq:ci_dev}
    S^d = \{s^{d_1}, s^{d_2}, ..., s^{d_{n^d}}\}
\end{equation}
\begin{equation}\label{eq:dev_specs}
    s^{d_i} = \{\tau^{d_i},\alpha^{d_i}\}, i = 1, 2, ..., n^d, \tau^{d_i} \in T
\end{equation}
\begin{equation}\label{eq:dev_actions}
    \alpha^{d_i} = \{a^{d_i}_1, a^{d_i}_2, ..., a^{d_i}_{m^{d_i}}\}
\end{equation}
\begin{multline}\label{eq:action_def_dev}
    a^{d_i}_j = \{\psi^{d_i}_j, \theta^{d_i}_j(\psi^{d_i}_j), \phi^{d_i}_j\}, 
    \psi^{d_i}_j \in \Psi(\tau^{d_i}),\\ \phi^{d_i}_j \in \mathbb{R}_{+}, j = 1, ..., m^{d_i}
\end{multline}
\begin{multline}\label{eq:action_param_dev}
\theta^{d_i}_j = 
\begin{cases}
    \{\gamma^{d_i}_j, e^{d_i}_j\} & \text{if $\gamma^{d_i}_j$ = access}\\
    \{\gamma^{d_i}_j, \zeta^{d_i}_j, \xi^{d_i}_j\} & \text{if $\gamma^{d_i}_j$ = exploit}\\
    \{\gamma^{d_i}_j, \pi^{d_i}_j\} & \text{if $\gamma^{d_i}_j$ = impact}
\end{cases}, \\
    \gamma^{d_i}_j \in \{\textrm{access}, \textrm{exploit}, \textrm{impact}\}
\end{multline}

Equations \eqref{eq:ci_link} to \eqref{eq:action_param_link} describe the specifications of network links as identified by the $l$ superscript. $s^{l_i}$, $\alpha^{l_i}$ and $a^{l_i}_j$ correspond respectively to the specification, list of possible actions and individual actions associated with the network link, in analogy to the device specifications. The same analogy applies to the number of network links ($n^l$) and number of possible actions ($m^{l_i}$). However, there are relevant differences between the specification of a network link and a device. $s^{l_i}$ is a set containing a pair of devices $\delta^{l_i}$ and a set of logical links $L^{l_i}$ besides $\alpha^{l_i}$. The pair $\delta^{l_i}$ represents the devices physically connected by the network link. The pairs $\lambda^{l_i}_k$ represent logical links associated with $l_i$ which connect the corresponding devices. There are $o^{l_i}$ logical links associated with $l_i$. The specification of possible actions is analogous to that of devices, as it also consists of a type ($\psi^{l_i}_j$), a list of parameters which are a function of this type ($\theta^{l_i}_j$) and a cost ($\phi^{l_i}_j$). However, the set of types associated with a network link is a function of the list of logical links ($\Psi(L^{l_i})$). Differences can also be noticed in the action parameters $\theta^{l_i}_j$ as category $\gamma^{l_i}_j$ can only assume values \textit{access} or \textit{impact}, actions of category \textit{access} can only be entry points, and actions of category \textit{impact} are associated with a specific logical link.

\begin{equation} \label{eq:ci_link}
S^l = \{s^{l_1}, s^{l_2}, ..., s^{l_{n^l}}\}
\end{equation}
\begin{equation} \label{eq:link_specs}
s^{l_i} = \{\delta^{l_i},\alpha^{l_i},L^{l_i}\}, i = 1, 2, ..., n^l
\end{equation}
\begin{equation} \label{eq:link_devs}
\delta^{l_i} = (d_{f^{l_i}}, d_{t^{l_i}})
\end{equation}
\begin{equation} \label{eq:link_actions}
\alpha^{l_i} = \{a^{l_i}_1, a^{l_i}_2, ..., a^{l_i}_{m^{l_i}}\}
\end{equation}
\begin{equation} \label{eq:logical_links}
L^{l_i} = \{\lambda^{l_i}_1, \lambda^{l_i}_2, ..., \lambda^{l_i}_{o^{l_i}}\}
\end{equation}
\begin{equation} \label{eq:logical_link_def}
\lambda^{l_i}_k = (d_{f^{l_i}_j}, d_{t^{l_i}_j}), k = 1, ..., o^{l_i}
\end{equation}
\begin{multline} \label{eq:action_def_link}
a^{l_i}_j = \{\psi^{l_i}_j, \theta^{l_i}_j(\psi^{l_i}_j), \phi^{l_i}_j\}, 
\psi^{l_i}_j \in \Psi(L^{l_i}),\\ \phi^{l_i}_j \in \mathbb{R}_{+}, j = 1, ..., m^{l_i}
\end{multline}
\begin{multline}\label{eq:action_param_link}
\theta^{l_i}_j = 
\begin{cases}
    \{\gamma^{l_i}_j, 1\} & \text{if $\gamma^{l_i}_j$ = access}\\
    \{\gamma^{l_i}_j, \lambda^{l_i}_k, \pi^{l_i}_j\}, \lambda^{l_i}_k \in L^{l_i}  & \text{if $\gamma^{l_i}_j$ = impact}
\end{cases}
,\\ \gamma^{l_i}_j \in \{\textrm{access}, \textrm{impact}\}
\end{multline}

For both devices and network links, the possible actions of the adversary are modeled as a discrete set. Therefore, any continuous quantity that could be manipulated as part of an attack must undergo some form of discretization before it can be integrated into the framework.

\subsubsection{CIE and Augmented SM State Transitions} \label{sec:cie}

The transitions of cyber-states $X_{cy}$ happen as a consequence of the actions performed by the adversary along the attack. The values of those states at a certain step in the attack define the corresponding impact to the industrial system at that point as well as the awareness and access the adversary can employ to take further steps. Equations \eqref{eq:X_cy} to \eqref{eq:M} describe the cyber-states. $X_d$, $X_l$ and $X_\alpha$ in the equations correspond respectively to state vectors associated with devices, links and actions. There is a one-to-one relation between each of those entities and a state. 

\begin{equation} \label{eq:X_cy}
X_{cy} = [X^{T}_{d}\;X^{T}_{l}\;X^{T}_{\alpha}]^{T}
\end{equation}
\begin{equation} \label{eq:X_d}
X_d = [x_{d_1}\;x_{d_2}\;...\;x_{d_{n^d}}]^{T}
\end{equation}
\begin{equation} \label{eq:X_l}
X_l = [x_{l_1}\;x_{l_2}\;...\;x_{l_{n^l}}]^{T}
\end{equation}
\begin{equation} \label{eq:X_alpha}
X_\alpha = [x_{a_1}\;x_{a_2}\;...\;x_{a_M}]^{T}
\end{equation}
\begin{equation} \label{eq:M}
M = \sum^{n^d}_{i=1}m^{d_i} + \sum^{n^l}_{i=1}m^{l_i}
\end{equation}

All states are discrete and they can assume a different set of values depending on the kind of entity they are associated with. Table \ref{tab:cyber_state_values} presents this association.

Tables \ref{tab:device_state_transitions} to \ref{tab:attack_state_transitions} present all possible state transitions for each entity type. On each table, the minus ($-$) and plus ($+$) superscripts indicate respectively the state before and after the corresponding transition. The symbols $\land$ and $\vert$ are employed respectively for \textit{and} and \textit{such that} when presenting the conditions for the transitions to occur.

\begin{table}
\caption{possible values which can be assumed by cyber-states for each entity type}
\label{tab:cyber_state_values}
\setlength{\tabcolsep}{3pt}
\begin{tabular}{p{50pt}p{50pt}p{115pt}}
\hline
\textbf{State}& 
\textbf{Entity Type}& 
\textbf{Possible Values} \\
\hline
$x_d$& 
device& 
 not visible, visible, accessible, compromised\\
\hline
$x_l$& 
link& 
 not accessible, accessible\\
\hline
$x_a$& 
action& 
 not accessible, accessible, active, invalid\\
\hline
\end{tabular}
\end{table}

\begin{table}
\caption{possible state transitions for device $d^{d_i}$ (state $x_{d^{d_i}})$}
\label{tab:device_state_transitions}
\setlength{\tabcolsep}{3pt}
\begin{tabular}{S{p{50pt}}S{p{115pt}}S{p{50pt}}}
\hline
${x^-_{d^{d_i}}}$& 
\textbf{Conditions}& 
${x^+_{d^{d_i}}}$ \\ 
\hline
not visible& 
$a^{d_{i'}}_j | \, ((i' \neq i) \,\land$ \newline $(\gamma^{d_{i'}}_j = \text{exploit}) \,\land \, (d^{d_i} \in \xi^{d_{i'}}_j))$& 
visible\\
\hline
not visible& 
$a^{l_{i'}}_j | \, ((\gamma^{l_{i'}}_j = \text{access})\,\land$ \newline \rule[-1ex]{0pt}{4ex} $(d^{d_i} \in \delta^{l_{i'}}))$ & 
visible\\
\hline
visible& 
$a^{d_{i}}_j \;| \; (\gamma^{l_{i}}_j = \text{access})$ & 
accessible\\
\hline
accessible& 
$a^{d_{i}}_j \;| \;(\gamma^{l_{i}}_j = \text{exploit})$ & 
compromised\\
\hline
\end{tabular}
\end{table}

\begin{table}
\caption{possible state transitions for link $l^{l_i}$ (state $x_{l^{l_i}})$}
\label{tab:link_state_transitions}
\setlength{\tabcolsep}{3pt}
\begin{tabular}{S{p{50pt}}S{p{115pt}}S{p{50pt}}}
\hline
${x^-_{l^{l_i}}}$& 
\textbf{Conditions}& 
${x^+_{l^{l_i}}}$ \\
\hline
not accessible& 
$a^{d_{i'}}_j | \, ((\gamma^{d_{i'}}_j = \text{exploit})\,\land$ \newline \rule[-1ex]{0pt}{4ex} $(d_{i'} \in \delta^{l_i})\,\land \, (\delta^{l_i} \cap \zeta^{d_{i'}}_j \neq \emptyset))$ & 
accessible\\
\hline
not accessible& 
$a^{l_{i}}_j \;| \;(\gamma^{l_{i}}_j = \text{access})$  & 
accessible\\
\hline
\end{tabular}
\end{table}

\begin{table}
\caption{possible state transitions for action $a_j$ (state $x_{a_j}$)}
\label{tab:attack_state_transitions}
\setlength{\tabcolsep}{3pt}
\begin{tabular}{S{p{50pt}}S{p{115pt}}S{p{50pt}}}
\hline
${x^-_{a_j}}$& 
\textbf{Conditions}& 
${x^+_{a_j}}$ \\
\hline
not accessible& 
$(x^+_{d^{d_i}} = \text{visible}) \,\land$ \newline $(a_j = a^{d_{i}}_{j'}) \,\land \, (\gamma^{d_i}_{j'} = \text{access})$  & 
accessible\\
\hline
not accessible& 
$(x^+_{d^{d_i}} = \text{accessible}) \,\land$ \newline $(a_j = a^{d_{i}}_{j'}) \,\land \, (\gamma^{d_i}_{j'} = \text{exploit})$  & 
accessible\\
\hline
not accessible& 
$(x^+_{l^{l_i}} = \text{accessible}) \,\land\, (a_j \in \alpha^{l_{i}}) $  & 
accessible\\
\hline
not accessible& 
$a^{d_{i}}_{j'} \,| \, ((\gamma^{d_{i}}_{j'} = \text{exploit}) \,\land$ \newline \rule[-1ex]{0pt}{4ex} $ (a_j \in \zeta^{d_{i}}_{j'}))$ & 
accessible\\
\hline
accessible& 
$a_j$  & 
active\\
\hline
accessible & 
$(x^+_{d^{d_i}} = \text{accessible}) \,\land$ \newline $(a_j = a^{d_{i}}_{j'}) \,\land \, (\gamma^{d_i}_{j'} = \text{access})$  & 
invalid\\
\hline
\end{tabular}
\end{table}

The CIE is responsible for keeping track of cyber-states and their transitions and extracting related information which is employed for system simulation and attack optimization. Equations \eqref{eq:cie_f} and \eqref{eq:cie_g} represent the computations performed by the CIE when the adversary performs action $a_j$. 

\begin{equation} \label{eq:cie_f}
X^{+}_{cy} = f_{cy}(X^{-}_{cy},a_j)
\end{equation}
\begin{equation} \label{eq:cie_g}
A^{+} = g_{cy}(X^{+}_{cy})
\end{equation}

Function $f_{cy}(.)$ corresponds to execution of state transitions as defined in tables \ref{tab:device_state_transitions} to \ref{tab:attack_state_transitions} based on actions taken by the adversary. 

Defining $S_\alpha$ as the set of all actions $a_j$, $j = 1, 2, ..., M$, where the state $x_{a_j}$ in $X_{cy}$ refers to $a_j$, function $g_{cy}(.)$ can then be defined as presented in \eqref{eq:cie_g_def}. 

\begin{equation} \label{eq:cie_g_def}
g_{cy}(X_{cy}) = \{a_j \in S_\alpha | x_{a_j} = \text{accessible}\}
\end{equation}

Both $f_{cy}(.)$ and $g_{cy}(.)$ are computed multiple times for each attack scenario, once for every action performed by the adversary. It is also a role of CIE to provide the sequence of cyber-states defining an attack scenario ($\overline{X}_{cy}$) for the SM so that the KPI can be calculated according to \eqref{eq:y_def} (Fig. \ref{fig:solution_diagram}). Equation \eqref{eq:f_bar} presents the computation of such sequence for scenario $\sigma$. This computation is performed once for each considered scenario. $\overline{a}^\sigma$ corresponds to the sequence of adversary actions performed during cyberattack scenario $\sigma$. 

\begin{equation} \label{eq:f_bar}
    \overline{X}^\sigma_{cy} = \overline{f}_{cy}(X^0_{cy},\overline{a}^\sigma)
\end{equation}
\begin{equation} \label{eq:a_bar}
    \overline{a}^\sigma = [a^\sigma_1, a^\sigma_2, ..., a^\sigma_{z^\sigma_{cy}}]
\end{equation}

Standard initial conditions for all cyber-states, as indicated by the superscript $0$, are defined in \eqref{eq:initial_conditions_device} to \eqref{eq:initial_conditions_attack} for each entity type. These definitions assume no devices or links are accessible or compromised at the beginning of the attacks, but they can be adjusted to fit any other situation.

\begin{equation} \label{eq:initial_conditions_device}
x^0_{d^{d_i}} = \text{not visible}
\end{equation}
\begin{equation} \label{eq:initial_conditions_link}
x^0_{l^{l_i}} = \text{not accessible}
\end{equation}
\begin{equation} \label{eq:initial_conditions_attack}
x^0_{a_j} = 
\begin{cases}
    \text{accessible,} & \text{if $\gamma_j$ = access $\land$ $e_j = 1$}\\
    \text{not accessible,} & \text{otherwise}
\end{cases}
\end{equation}

\subsection{Optimization Problem Formulation} \label{sec:optimization_problem}
The cyberattack optimization problem can be defined as obtaining the optimal sequence of actions $\overline{a}^{*}$ that minimizes the system KPI of interest ($y$) as presented in \eqref{eq:optimization_problem}. Defining $\beta$ as the attack budget available to the adversary, the optimization is also subject to constraint \eqref{eq:constraint}.

\begin{equation} \label{eq:optimization_problem}
    \overline{a}^{*} = \argmin_{\overline{a}}(y)
\end{equation}
\begin{equation} \label{eq:constraint}
    \sum^{z_{cy}}_{\kappa = 1}\phi_\kappa \leq \beta
\end{equation}

\subsection{AI-based Sequential Decision-Making Optimization} \label{sec:SDMO}
A SDMO engine is employed for pursuing the sequence of actions which is optimal in terms of causing the largest disruption to the industrial system as measured by the operational KPI of interest \eqref{eq:optimization_problem}. The SDMO interacts with the CIE (Fig. \ref{fig:solution_diagram}) multiple times per attack scenario, defining actions taken by the adversary at the current step and obtaining in response all possible actions that can be performed in the following step of the attack \eqref{eq:cie_g}. The SDMO also interacts with the SM in order to obtain KPI values corresponding to the current attack scenario \eqref{eq:y_def}. 

Monte Carlo Tree Search (MCTS) \cite{Browne2012} is an AI method which can be employed for solving such type of sequential decision making problem. It has been successfully applied in various similar contexts in recent years especially for games \cite{swiechowski2023}. It is, for instance, the main method behind AlphaGo – the first computer
program to defeat a professional human Go player \cite{silver2016}.

MCTS is a best-first search algorithm where a tree is iteratively built based on random exploration of sequences of states. Each node in the tree represents a point in the state space and child nodes are obtained by means of state transitions. Each node is associated with a score that is updated based on the rewards obtained while traversing the tree. This score is usually based on the UCT (upper confidence bound applied to trees) \cite{Browne2012} criteria which provides simple means for balancing exploration and exploitation. The search is performed based on multiple iterations. In each MCTS iteration the node with the highest score is chosen and expanded by means of a state transition. From the expanded node, a sequence of random state transitions is applied until reaching a terminal state, in a process referred to as rollout. The scores linked to each node in the path taken from the tree root to this terminal state are updated based on the rewards obtained during the rollout. The search ends when reaching a pre-defined limit in terms of number of iterations or computation time. At this point, the next state transition is defined based, for instance, on the choice that leads to the node with the highest score. The whole search process is repeated for selection of each additional state transition until such transition leads to a terminal state.

In the context of this work, states correspond to the cyber-state $X_{cy}$ \eqref{eq:X_cy}, transitions result from adversary actions \eqref{eq:cie_f} and the rewards are a function of the KPI $y$ which is calculated using the SM \eqref{eq:y_def}. Each path from the root to a terminal node corresponds to a sequence of cyber-states $\overline{X}_{cy}$ that define an attack scenario as described in section \ref{sec:cie}. Actions are chosen among the options defined by the set $A$ after its update based on \eqref{eq:cie_g}, and terminal states are achieved when the attack budget left after subtracting the costs of all previous adversary actions is not sufficient for an additional action, as indicated in constraint \eqref{eq:constraint}.  

MCTS has some characteristics which make it especially suited for solving the SDMO problem in the context defined here. It is a so-called anytime algorithm as it can produce a valid solution even if interrupted during computations. The SM can be employed as a black-box model that provides the reward corresponding to a certain cyber-state. Besides, rewards only need to be calculated for terminal nodes, which is beneficial due to the computational cost associated with the SM execution.

\section{Experiments and Results} \label{sec:experiments}
The electric power system, which is undergoing a transformation due to the proliferation of renewable generation sources and IoT devices, is becoming an increasingly popular target of cyber attacks.  As the devices such as rooftop solar systems, behind the meter storage (i.e. batteries), electric vehicles, and smart appliances have inherent electronic communications capabilities, the cyberattack surface of modern electric power grids has vastly expanded \cite{sahoo2019cyber}.  Therefore, power systems correspond to a very adequate domain for application of the proposed methodology.

For this purpose the PyCIGAR tool \cite{roberts2020} was employed. PyCIGAR\footnote{https://secpriv.lbl.gov/project/rmt-magic/} is a library based in Python programming language which wraps power system simulation models developed in formats such as OpenDSS\footnote{https://www.epri.com/pages/sa/opendss} and facilitates systematic testing and integration of external logic to the simulation, especially in terms of device control. As part of the work described here, the library was extended to include control and possible adversary actions for additional devices as described in section \ref{sec:attack_modeling}.

\subsection{Power Grid Modeling, KPIs, and Attacks}
\subsubsection{Power Grid Modeling}
\begin{figure} [h]
    \centering
    \includegraphics[scale=0.68]{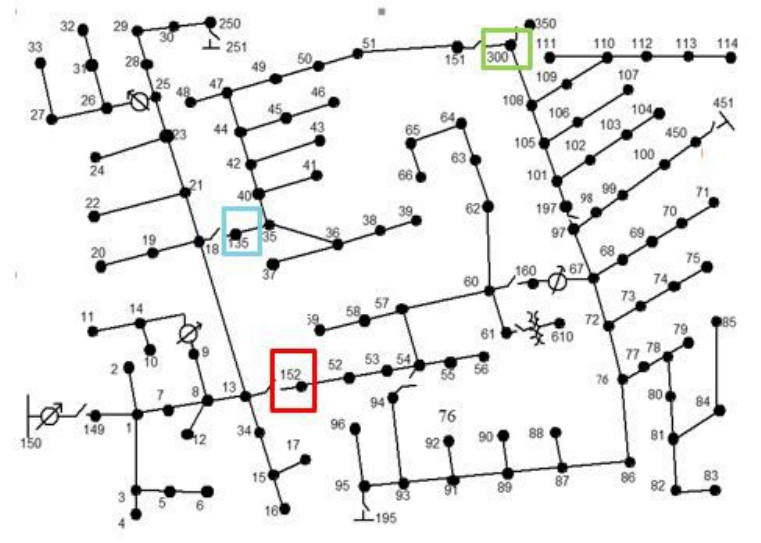}
    \caption{Shown here is the standard IEEE $123$-bus feeder model (for a single feeder) connected to the substation bus $150$. The $246$-bus network was built by mirroring the standard IEEE $123$-bus feeder and adding it to the existing substation, i.e., the power network consisted of one substation with two $123$-bus feeder models, henceforth referred to as Feeder $1$ and Feeder $2$. Boxes in the figure show the locations of normally open switches between the same buses ($135$, $152$, and $300$) in Feeder $1$ and $2$.}
    \label{fig:IEEE123}
\end{figure}

For the experiments, the standard IEEE $123$-bus feeder model \cite{WHK} was modified to custom-build a $246$-bus power distribution network in OpenDSS as described in Fig.~\ref{fig:IEEE123}. 

The $246$-bus network model was augmented by normally open switches to enable  topology reconfiguration operations. The standard IEEE $123$-bus feeder model was also modified to include additional PV (photovoltaic inverters) on every bus to represent a distribution network with high penetration of renewable energy generation. PV and load profiles consisted of a four-hour time window representing a sample operation from $9\!:\!00$ a.m. to $1\!:\!00$ p.m. with a $1s$ sampling rate. The PV profiles were chosen to allow inter-feeder topology reconfiguration over the entire time series without overloading the substation transformer, i.e., if the topology is reconfigured to bring Feeder $2$ onto Feeder $1$, then the substation becomes heavily loaded, but this does not cause overloading or severe under voltage issues.

Three battery energy storage systems (BESS) were added to capture a realistic network with storage devices. They were randomly dispersed in the network on buses $33$, $48$, and $71$ in Feeders $1$ and $2$. The control logic was set up to perform peak shaving and valley filling functions from a centralized controller that monitors the substation active power in-feed. The valley filling function enables charging when the net active power at the substation is below a specified threshold. Similarly, the peak shaving function discharges the BESS when the net active power at the substation exceeds a specified threshold. In between the thresholds for peak shaving and valley filling the BESS is neither charging nor discharging.

\subsubsection{KPIs} \label{sec:KPIs}
Key performance indicators (KPIs) were defined to quantify how well the system is operating, which can then be used for measuring efficacy of an attack in disrupting operation \eqref{eq:y_def}. The KPI employed for the analysis is a combination of voltage imbalance ($V\!I$) and substation power factor ($S\!P\!F$). We refer to this KPI as $V\!I\!S\!P\!F$ hereafter. The corresponding definitions are presented below. 

Consider a power network comprising of $N$ nodes modeled as a connected graph $\mcG = (\mcV, \mcE)$, whose nodes $\mcV =
\{1,\ldots,N\}$ correspond to buses, and edges $\mcE$ to undirected lines. Then $V\!I$ is defined as:
\begin{equation} \label{eq:VI}
     V\!I=\max_{i_{ph} \in \{a_{ph},b_{ph},c_{ph}\}, j_\mcV \in \mcV} \left\{\frac{|V_{i_{ph},j_\mcV}-\hat{V}_{j_\mcV}|}{\hat{V}_{j_\mcV}}\right\}
\end{equation}
where $V_{i_{ph},j_\mcV}$ are the line-to-neutral voltage magnitudes of electrical phase $i_{ph} \in \{a_{ph},b_{ph},c_{ph}\}$ on bus $j_\mcV \in \mcV$ and $\hat{V}_{j_\mcV} = \sum_{i_{ph}} V_{i_{ph},j_\mcV}/3$. Similarly, $S\!P\!F$ is defined as:
\begin{equation} \label{eq:SPF}
S\!P\!F=\left|\cos\left(\tanh\left(\frac{q_{ss}}{p_{ss}}\right)\right)\right|
\end{equation}
where $q_{ss}$ and $p_{ss}$ are respectively the net reactive and active power at the substation obtained from summation of the values from the three phases. Lastly, the combined $V\!I\!S\!P\!F$ KPI is defined as:
\begin{equation}
   V\!I\!S\!P\!F=\frac{1-V\!I+S\!P\!F}{2}.
\end{equation}
The KPI values vary on a range between 0 to 1. Attacks that minimize this KPI can be considered the most effective \eqref{eq:optimization_problem}.

\subsubsection{Attack Effects in the Power System} \label{sec:attack_modeling}
The PyCIGAR framework provides means to facilitate modeling of power grid devices and controllers. In the library, device models represent the interaction of specific elements with the grid, whereas controller models, as the name states, are employed to define the corresponding control logic. Possible adversary actions were modeled as part of the controller models. This modeling results in desirable characteristics such as the possibility of starting or ending any of those actions at any point during the simulation and evaluating the combined effects of multiple actions.

Table \ref{tab:device_attacks} presents an overview of the modeled devices and the associated possibilities in terms of adversary actions. All these actions correspond to category \textit{impact} as described in section \ref{sec:actioncategories}.
\begin{table*}[!ht] 
\caption{Devices modeled in PyCIGAR and corresponding possible adversary actions}
\label{tab:device_attacks}
  \centering
\begin{tabular}{ l l l }
\hline

 \textbf{Device} & \textbf{Adversary Action} & \textbf{Description} \\
 \hline
 \multirow {3}{7em}{PV Inverter Device} & Connect/Disconnect & Disconnects the PV Device for the specified time \\
	& Voltage Break Points & Modifies the Volt/Var and Volt/Watt behavior \\
	& Unbalanced & Creates an unbalanced output across the 3 phases \\
 \hline
\multirow {4}{7em}{Battery Storage Device} & Operation Mode & Modifies the operation mode (e.g.: charge instead of discharge) \\
	& Power Injection & Forces to discharge the battery with maximum power  \\
	& Power Consumption & Forces to charge the battery with maximum power  \\
	& Battery Settings & Modifies the control parameters (e.g.: reduce the max ramp rate) \\
\hline
Switch Device & Open/Close (Topology) & Operates two connected switches to change the topology\\
\hline
Capacitor Device & Curtailment & Reduces the capacity of the capacitor bank  \\
\hline
\multirow {4}{7em}{Regulator Device} & Change Settings & Modifies the regulator settings\\
	& Regulator Deactivate & Fixes the regulator to a specific tap \\
	& Regulator Prohibit Control & Prohibits the regulator from executing controls \\
	& Change Taps & Changes the regulators tap     \\
\hline
\label{table:5} 
\end{tabular}
\end{table*}

\subsection{System Model Augmentation}

\begin{figure}[h]
    \centering
    \includegraphics[scale=0.135]{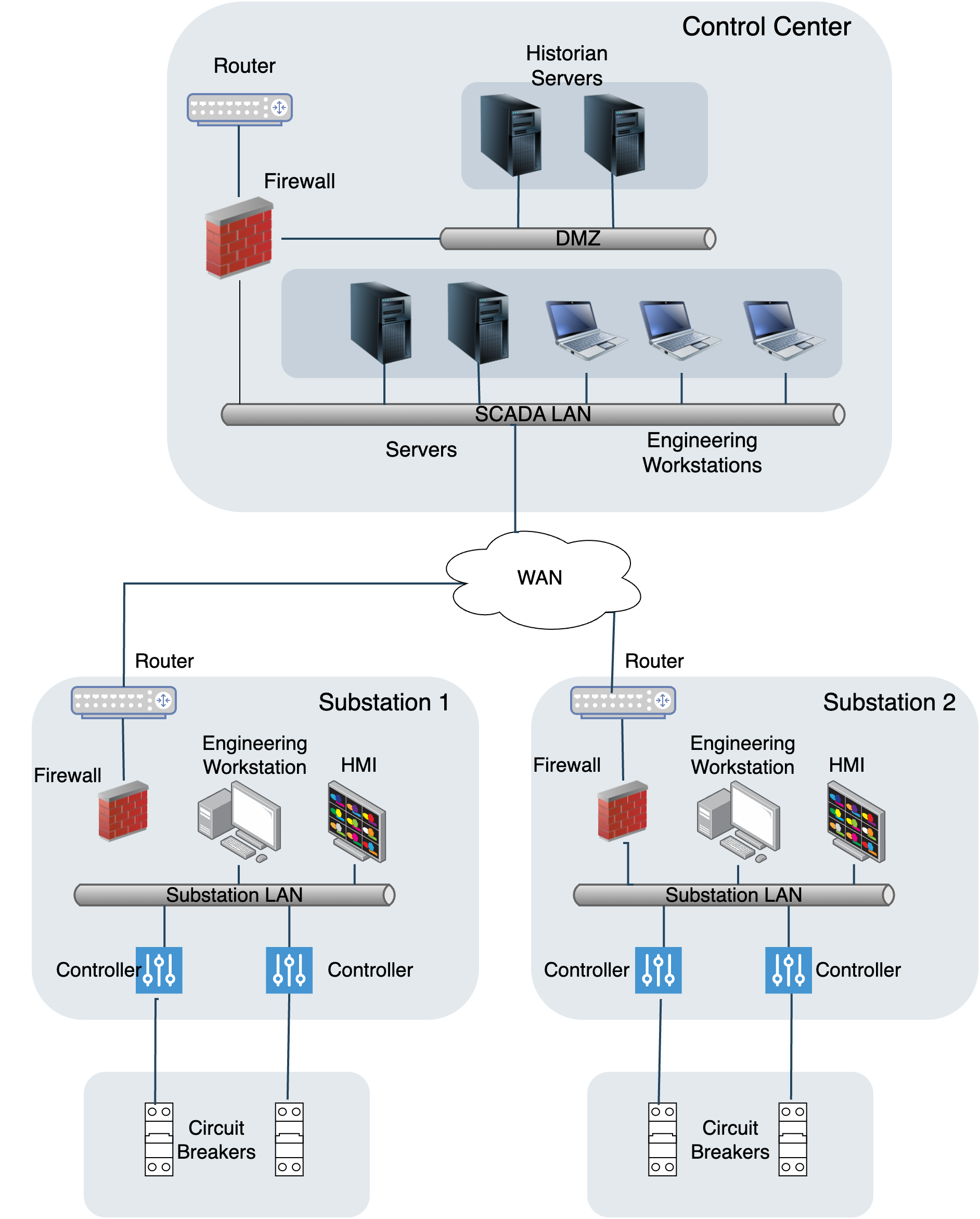}
    \caption{System architecture pattern employed for definition of the communication network.}
    \label{fig:Network_diag}
\end{figure}

The communication network defined for the experiments was based on a distribution feeder automation design guide \cite{Cisco_guide}, providing means for the creation of realistic scenarios for assessing the performance of the proposed methodology. Various network devices, including routers, switches, and device controllers, were included and associated with the power system of interest by following the SCADA architecture pattern presented in Fig. \ref{fig:Network_diag}. Table \ref{tab:devices} lists the different categories of devices included in the computer network and the number of devices for each category.

\begin{table}[ht]
  \centering
  \caption{Devices included in the network architecture }
  \label{tab:devices}
  \begin{tabular}{lc}
    \hline
    \textbf{Type of Device} & \textbf{Number of Devices} \\
    \hline
    Router (Edge Device) & 47  \\
    Firewall & 2  \\
    Workstation & 5\\
    BESS  & 3 \\
    Capacitor Bank & 4 \\
    Switch Controller & 1  \\
    Load Controller & 91  \\
    PV Inverters & 91  \\
    \hline
  \end{tabular}
\end{table}

Communication network devices and topology were encoded using the NetJSON\footnote{https://netjson.org/} format. NetJSON is a data interchange format based on JavaScript Object Notation (JSON) specifically designed to describe the fundamental components of layer 2 and layer 3 networking. It was chosen as it offers a standardized and structured approach to represent network-related information, including device configurations, monitoring data, network topology, and routing information.

The mapping of all logical links corresponding to each network link was automated. The functionality of the Python package NetworkX\footnote{https://networkx.org/} was employed to traverse the computer network graph to determine such logical links. It was assumed that communication is performed between each controller and the control center. All the source nodes were first identified, and then the graph was traversed to identify the sink nodes and the paths connecting sources to sinks. The result, in the form of a list of logical links associated with each physical link, was stored using the NetJSON format.

Details regarding potential attacker actions associated with each device and network link within different categories were also encoded using NetJSON, following the definitions presented in section \ref{sec:actioncategories}. For each action, the primary information recorded consisted of its category and its cost. 

As mentioned above, the expertise required for implementing the proposed methodology is similar to that required for performing standard cybersecurity risk assessments from an IT perspective, with the addition of OT support. For the experiments, a group of experts in the fields of penetration testing, OT cybersecurity, and power systems performed the analysis that produced the cost values employed for the experiments. After grouping the actions in each category into a reduced number of cost groups, the experts considered that the direct definition of the costs was feasible and application of the Elo rating system was not required. One example of cost group definition is provided here for illustration. Actions of category \textit{impact} associated with devices were grouped in two cost levels: if the functionality corresponding to the action was already present in the device, such as changing a configuration, it was provided with a lower cost, while in case more complex manipulation was required, such as code injection, a higher cost was provided. 

The Delphi Technique described in section \ref{sec:costsquantification} was also employed to facilitate consensus-building among the experts. The process involved three rounds of assessment and feedback, allowing the experts to refine their judgments and reach greater agreement on the relative costs associated with various attack actions. 

Here is an example of how the Delphi Technique was employed. We analyzed the technique "Exploit Public-Facing Application" (T0819) \footnote{https://attack.mitre.org/techniques/T0819/}, which was associated with a set of possible attacker actions. Each expert had to provide cost values ranging from 0 to 20.  Initial estimates varied from 12 -- 15. Experts considered factors like the prevalence of software vulnerabilities, the accessibility of exploitation tools, and the attacker's skill level. To assign the scores, a qualitative research approach was followed by creating a generic threat matrix inspired by Woodered et. al \cite{woodard2007categorizing}

In the subsequent rounds, experts were provided with anonymized feedback summarizing the range and rationale behind previous estimates. Discussions highlighted the balance between easy vulnerability discovery and the targeted effort needed for zero-day exploits, as well as the role of exploit frameworks in simplifying the attack process. The Delphi process encouraged experts to reevaluate their assumptions, leading to greater cost estimate convergence, finally settling on a value of 14. This reflects the significant resources and expertise attackers would likely need to execute a successful exploit against a public-facing ICS application. 

\subsection{CIE Implementation and Optimization Results}
CIE logic for updating and extracting information from the cyber-states as described in section \ref{sec:cie} was implemented based on the Julia programming language, using the POMDP library\footnote{https://github.com/JuliaPOMDP/POMDPs.jl}. This implementation included integration with the PyCIGAR/OpenDSS-based SM for calculation of rewards. The corresponding MCTS solver\footnote{https://github.com/JuliaPOMDP/MCTS.jl} was employed for solution of the sequential decision making optimization as described in section \ref{sec:SDMO}.

As a baseline for assessing the performance of the methodology, a large number of attack scenario samples were randomly generated and evaluated for computation of the corresponding reward \eqref{eq:y_def} using a node from a super computing cluster comprising an Intel Xeon Gold 5218 processor with 32 cores and 1584GB of RAM. CIE logic, including state transitions and attack budget constraints, was enforced in order to define attack scenarios the same way as in the search. All cores of the computing node were employed to generate the samples and memory sharing mechanisms were applied to exchange information about the scenarios explored by each process in order to ensure the diversity of the generated attack scenario samples. A total of 504 wall-clock hours were employed for such processing, producing 504217 attack scenario samples and the corresponding reward values. An empirical cumulative distribution function (ECDF) based on those reward values was used to evaluate the performance of the proposed SDMO methodology, indicating the percentage of random samples that produce rewards which are lower then the one resulting from optimization. This percentage will be hereafter referred to as $p_{CDF}$. The higher this value, the more effective the corresponding attack scenario is in disrupting operation, using the random samples as reference.

SDMO experiments have been performed on a computer running Ubuntu 18.04 operating system, comprising Intel Xeon Silver 4210 processor and 187GB of RAM. Table \ref{tab:results} presents information about each test in terms of number of iterations and results concerning $p_{CDF}$ and total duration combining all search steps until reaching a terminal state. Computations in this case involve a single CPU processor. All tests were performed considering the same attack budget and MCTS configurations other than number of iterations, which was varied as shown in the table. The duration values presented in the table were normalized based on a fixed iteration duration for a fair comparison, as iterations performed later benefited from stored results produced by previous ones.

It can be noticed that all tests resulted in high $p_{CDF}$ values. Most results are within 0.951 and 0.984 with one exception in terms of lower value (0.931 in scenario \#6) and another corresponding to a high value ($>$0.999 in scenario \#4). It can also be noticed from the results that increasing the number of iterations has an expected impact in duration but it does not seem to have a relevant impact in $p_{CDF}$. One hypothesis that could explain this behavior is that all values tested for the number of iterations are the same order of magnitude and it would require order of magnitude increases in this number to obtain systematic improvements. Such order of magnitude increase would be feasible by applying some approaches described in section \ref{sec:future_work}.

\begin{table}[ht]
\centering
\caption{Optimization experiments and results}
\label{tab:results}
\begin{tabular}{cccc}
\hline
\textbf{Test\#}& 
\textbf{Iterations}& 
\textbf{Duration [h]}&
$p_{CDF}$\\
\hline
1& 
2000& 
27&
0.984\\

2& 
2000& 
24&
0.962\\

3& 
2000& 
21&
$>$0.999\\

4& 
5000& 
60&
0.951\\

5& 
5000& 
69&
0.984\\

6& 
5000& 
60&
0.931\\

7& 
7500& 
103&
0.981\\

8& 
7500& 
64&
0.962\\

9& 
7500& 
103&
0.984\\

\hline
\end{tabular}
\end{table}

As an illustration of the adversary actions associated with those attack scenarios, below is a description of the steps yielded in scenario \#3 which corresponds to the best $p_{CDF}$ in table \ref{tab:results}:
\begin{enumerate}
    \item The adversary uses as entry point (\textit{access} category) a network link connecting an edge device that concentrates communication from multiple controllers to a router that connects  many of such edge devices to a substation (Figure \ref{fig:Network_diag}). This network link is exposed in the field and not behind a firewall. This can be considered a good trade off in terms of attack cost and impact.
    \item Once the link is accessible, the adversary performs false data injection attacks (\textit{impact} category) affecting four controllers corresponding to three controllable loads and one PV inverter.
\end{enumerate}

The network link used as entry point in this scenario and the devices connected by it may represent relevant risks to be addressed. This is also confirmed by the baseline random search results, as the most impactful attack scenario obtained in this case was based on the same router that connects to a substation. There are other routers and network links in the system which are similar in terms of device characteristics, function and connectivity. The higher relevance of these devices/link compared to other ones could only be identified when taking the power system operation into account.

The analysis of the other 8 cases in table \ref{tab:results} also provides valuable insights. The entry point in all of those cases was a load controller, and attack scenarios included manipulation of loads. In some cases, network-based attacks affecting PV inverters were also performed. One interesting aspect to consider is that in 5 out of those 8 scenarios the same load controller - out of 91 load controllers and hundreds of other devices and network links - was employed as entry point. This provides evidence that attacks targeting this load controller as entry point can be a relevant threat, even though, in terms of device characteristics and computer network connectivity, it is similar to the other 90 load controllers. Similarly to the analysis of scenario \#3 above, the relevance of this specific controller could only be identified by explicitly assessing the impact of the attacks in the power grid as enabled by the proposed methodology.

Results can also be evaluated in terms of how the budget was spent by the adversary in terms of action categories. Considering the attack scenarios that correspond to the results in table \ref{tab:results}, it was spent as follows:
\begin{itemize}
    \item \textit{access} category (entry points): 44\%
    \item \textit{access} category (lateral movement): 8\%
    \item \textit{exploit} category: 10\%
    \item \textit{impact} category: 38\%
\end{itemize}

The budget was spent mostly on entry points and \textit{impact} category actions. This points to a strategy where the adversary first chooses a device/link corresponding to a good trade off between attack cost and impact, even if corresponding to a large percentage of its budget. Chosen targets are easily exploitable, e.g. associated with known and easy to implement exploits, and once they are exploited the adversary focuses in disturbing as much as possible the system operation from that location by means of \textit{impact} category actions. 

\section{Conclusion} \label{sec:conclusion}

This paper presented a methodology for identifying the most critical risks affecting an industrial system in terms of potential cyberattack scenarios. Criticallity is assessed in terms of impact in operational performance by means of a defined KPI. Therefore, the proposed framework performs a cybersecurity assessment with focus on the operator's point-of-view. The identification of the most critical cyberattacks is pursued by means of an AI-based tree search algorithm (MCTS), employed to perform a sequential decision-making optimization task. 

The methodology leverages simulation models (SMs) developed using standard engineering tools which are augmented to include information about the computer network topology and devices and potential adversary actions. An attack cost-budget mechanism is employed to define adjustable constraints to the adversary's actions. All the information required for augmenting the SM is represented using the NetJSON standard. Once the augmented SM is prepared, the methodology provides means for tracking cyber-states that describe the progression of attack scenarios, consisting of a sequence of adversary actions, in a way that enables the use of optimization/search methods for identification of the most critical possibilities. 

Experiments provided evidence of the successful application of the methodology based on a power distribution system use case. The results of an extensive random search performed using super computing infrastructure were used as baseline for comparison. Limitations to be addressed and opportunities for further development are discussed below.

\subsection{Limitation and Future Work} \label{sec:future_work}

One of the key opportunities for improvement of the proposed methodology is the computational performance of the SDMO solution, as it relies on the SM for obtaining KPI values. Considering the MCTS solution, one clear possibility to explore consists of parallelizing the calculations based on methods such as the one proposed by Liu et al. \cite{Liu2020}. This kind of enhancement can be very desirable as it may provide benefits which are application agnostic. Heuristics can also be designed to speedup MCTS computations \cite{Browne2012}.
An additional possibility for reducing computation costs is the use of surrogate models to replace the original SM \cite{bhosekar2018}, resembling the strategy used in AlphaGo \cite{silver2016} where the MCTS method was combined with machine learning models for computationally efficient simulation and definition of next action. The use of more recent methods based on evolutions of those approaches can also be explored such as Gumbel MuZero \cite{danihelka2022policy}, including associated computationally efficient implementations \cite{deepmind2020jax}.

Another important topic for future work is the integration into the methodology of additional information, methods and tools available in the OT cybersecurity literature and standard practices, especially considering the definition of potential adversary actions, costs and budgets. Further incorporation of extensively employed cybersecurity resources such as MITRE ATT\&CK and CVSS in such definitions can potentially enhance the methodology and promote its acceptance and adoption. Relevant aspects to explore in this context include alternative definitions of what the costs represent and the use of parameterized means for their quantification to account for dependencies on other actions or system changes, e.g. modifications in cyber defenses. Functionalities for integrating information from tools such as network scanners can also greatly facilitate the collection of information required to implement the proposed methodology.

\section{Acknowledgement}

Any opinions, findings, conclusions, or recommendations
expressed in this material are those of the authors and do not
necessarily reflect those of the sponsors of this work.

\bibliographystyle{IEEEtran}
\bibliography{references}

\begin{thebibliography}{10}
\providecommand{\url}[1]{#1}
\csname url@samestyle\endcsname
\providecommand{\newblock}{\relax}
\providecommand{\bibinfo}[2]{#2}
\providecommand{\BIBentrySTDinterwordspacing}{\spaceskip=0pt\relax}
\providecommand{\BIBentryALTinterwordstretchfactor}{4}
\providecommand{\BIBentryALTinterwordspacing}{\spaceskip=\fontdimen2\font plus
\BIBentryALTinterwordstretchfactor\fontdimen3\font minus \fontdimen4\font\relax}
\providecommand{\BIBforeignlanguage}[2]{{%
\expandafter\ifx\csname l@#1\endcsname\relax
\typeout{** WARNING: IEEEtran.bst: No hyphenation pattern has been}%
\typeout{** loaded for the language `#1'. Using the pattern for}%
\typeout{** the default language instead.}%
\else
\language=\csname l@#1\endcsname
\fi
#2}}
\providecommand{\BIBdecl}{\relax}
\BIBdecl

\bibitem{corallo2022}
\BIBentryALTinterwordspacing
A.~Corallo, M.~Lazoi, M.~Lezzi, and A.~Luperto, ``Cybersecurity awareness in the context of the industrial internet of things: A systematic literature review,'' \emph{Computers in Industry}, vol. 137, p. 103614, 2022. [Online]. Available: \url{https://www.sciencedirect.com/science/article/pii/S0166361522000094}
\BIBentrySTDinterwordspacing

\bibitem{LeBlanc2017}
K.~Le~Blanc, A.~Ashok, L.~Franklin, J.~Scholtz, E.~Andersen, and M.~Cassiadoro, ``Characterizing cyber tools for monitoring power grid systems: What information is available and who needs it?'' in \emph{2017 IEEE International Conference on Systems, Man, and Cybernetics (SMC)}, 2017, pp. 3451--3456.

\bibitem{review_yohanandhan2020cyber}
R.~V. Yohanandhan, R.~M. Elavarasan, P.~Manoharan, and L.~Mihet-Popa, ``Cyber-physical power system (cpps): A review on modeling, simulation, and analysis with cyber security applications,'' \emph{IEEE Access}, vol.~8, pp. 151\,019--151\,064, 2020.

\bibitem{attack_graph_zhang2015power}
Y.~Zhang, L.~Wang, Y.~Xiang, and C.-W. Ten, ``Power system reliability evaluation with scada cybersecurity considerations,'' \emph{IEEE Transactions on Smart Grid}, vol.~6, no.~4, pp. 1707--1721, 2015.

\bibitem{attack_graph_8720257}
A.~T. Al~Ghazo, M.~Ibrahim, H.~Ren, and R.~Kumar, ``A2g2v: Automatic attack graph generation and visualization and its applications to computer and scada networks,'' \emph{IEEE Transactions on Systems, Man, and Cybernetics: Systems}, vol.~50, no.~10, pp. 3488--3498, 2020.

\bibitem{attack_graph_stefanov2015scada}
A.~Stefanov, C.-C. Liu, M.~Govindarasu, and S.-S. Wu, ``Scada modeling for performance and vulnerability assessment of integrated cyber--physical systems,'' \emph{International Transactions on Electrical Energy Systems}, vol.~25, no.~3, pp. 498--519, 2015.

\bibitem{cao2018assessing}
C.~Cao, L.-P. Yuan, A.~Singhal, P.~Liu, X.~Sun, and S.~Zhu, ``Assessing attack impact on business processes by interconnecting attack graphs and entity dependency graphs,'' in \emph{Data and Applications Security and Privacy XXXII: 32nd Annual IFIP WG 11.3 Conference, DBSec 2018, Bergamo, Italy, July 16--18, 2018, Proceedings 32}.\hskip 1em plus 0.5em minus 0.4em\relax Springer, 2018, pp. 330--348.

\bibitem{9322321}
M.~A. Haque, S.~Shetty, C.~A. Kamhoua, and K.~Gold, ``Integrating mission-centric impact assessment to operational resiliency in cyber-physical systems,'' in \emph{GLOBECOM 2020 - 2020 IEEE Global Communications Conference}, 2020, pp. 1--7.

\bibitem{ou2005mulval}
X.~Ou, S.~Govindavajhala, A.~W. Appel \emph{et~al.}, ``Mulval: A logic-based network security analyzer.'' in \emph{USENIX security symposium}, vol.~8.\hskip 1em plus 0.5em minus 0.4em\relax Baltimore, MD, 2005, pp. 113--128.

\bibitem{ou2006scalable}
X.~Ou, W.~F. Boyer, and M.~A. McQueen, ``A scalable approach to attack graph generation,'' in \emph{Proceedings of the 13th ACM conference on Computer and communications security}, 2006, pp. 336--345.

\bibitem{tayouri2023survey}
D.~Tayouri, N.~Baum, A.~Shabtai, and R.~Puzis, ``A survey of mulval extensions and their attack scenarios coverage,'' \emph{IEEE Access}, 2023.

\bibitem{patapanchala2016exploring}
P.~S. Patapanchala, C.~Huo, R.~B. Bobba, and E.~Cotilla-Sanchez, ``Exploring security metrics for electric grid infrastructures leveraging attack graphs,'' in \emph{2016 IEEE Conference on Technologies for Sustainability (SusTech)}.\hskip 1em plus 0.5em minus 0.4em\relax IEEE, 2016, pp. 89--95.

\bibitem{kriaa2015survey}
S.~Kriaa, L.~Pietre-Cambacedes, M.~Bouissou, and Y.~Halgand, ``A survey of approaches combining safety and security for industrial control systems,'' \emph{Reliability engineering \& system safety}, vol. 139, pp. 156--178, 2015.

\bibitem{datalog_ceri1989you}
S.~Ceri, G.~Gottlob, L.~Tanca \emph{et~al.}, ``What you always wanted to know about datalog(and never dared to ask),'' \emph{IEEE transactions on knowledge and data engineering}, vol.~1, no.~1, pp. 146--166, 1989.

\bibitem{semertzis2022quantitative}
I.~Semertzis, V.~S. Rajkumar, A.~{\c{S}}tefanov, F.~Fransen, and P.~Palensky, ``Quantitative risk assessment of cyber attacks on cyber-physical systems using attack graphs,'' in \emph{2022 10th Workshop on Modelling and Simulation of Cyber-Physical Energy Systems (MSCPES)}.\hskip 1em plus 0.5em minus 0.4em\relax IEEE, 2022, pp. 1--6.

\bibitem{mitre_attack}
MITRE, ``{MITRE ATT\&CK},'' \emph{URL: https://attack.mitre.org}, 2021.

\bibitem{hasan2018vulnerability}
S.~Hasan, A.~Ghafouri, A.~Dubey, G.~Karsai, and X.~Koutsoukos, ``Vulnerability analysis of power systems based on cyber-attack and defense models,'' in \emph{2018 IEEE Power \& Energy Society Innovative Smart Grid Technologies Conference (ISGT)}.\hskip 1em plus 0.5em minus 0.4em\relax IEEE, 2018, pp. 1--5.

\bibitem{Hasan2020}
\BIBentryALTinterwordspacing
S.~Hasan, A.~Dubey, G.~Karsai, and X.~Koutsoukos, ``A game-theoretic approach for power systems defense against dynamic cyber-attacks,'' \emph{International Journal of Electrical Power \& Energy Systems}, vol. 115, 2020. [Online]. Available: \url{http://www.sciencedirect.com/science/article/pii/S0142061519302807}
\BIBentrySTDinterwordspacing

\bibitem{okoli2004delphi}
C.~Okoli and S.~D. Pawlowski, ``The delphi method as a research tool: an example, design considerations and applications,'' \emph{Information \& management}, vol.~42, no.~1, pp. 15--29, 2004.

\bibitem{glickman1999}
M.~E. Glickman and A.~C. Jones, ``Rating the chess rating system,'' \emph{Chance}, vol.~12, no.~2, pp. 21--28, 1999.

\bibitem{zheng2023judging}
L.~Zheng, W.-L. Chiang, Y.~Sheng, S.~Zhuang, Z.~Wu, Y.~Zhuang, Z.~Lin, Z.~Li, D.~Li, E.~P. Xing, H.~Zhang, J.~E. Gonzalez, and I.~Stoica, ``Judging llm-as-a-judge with mt-bench and chatbot arena,'' 2023.

\bibitem{Browne2012}
C.~B. Browne, E.~Powley, D.~Whitehouse, S.~M. Lucas, P.~I. Cowling, P.~Rohlfshagen, S.~Tavener, D.~Perez, S.~Samothrakis, and S.~Colton, ``A survey of monte carlo tree search methods,'' \emph{IEEE Transactions on Computational Intelligence and AI in Games}, vol.~4, no.~1, pp. 1--43, 2012.

\bibitem{swiechowski2023}
M.~{\'S}wiechowski, K.~Godlewski, B.~Sawicki, and J.~Ma{\'n}dziuk, ``Monte carlo tree search: A review of recent modifications and applications,'' \emph{Artificial Intelligence Review}, vol.~56, no.~3, pp. 2497--2562, 2023.

\bibitem{silver2016}
D.~Silver, A.~Huang, C.~J. Maddison, A.~Guez, L.~Sifre, G.~van~den Driessche, J.~Schrittwieser, I.~Antonoglou, V.~Panneershelvam, M.~Lanctot, S.~Dieleman, D.~Grewe, J.~Nham, N.~Kalchbrenner, I.~Sutskever, T.~Lillicrap, M.~Leach, K.~Kavukcuoglu, T.~Graepel, and D.~Hassabis, ``Mastering the game of go with deep neural networks and tree search,'' \emph{Nature}, vol. 529, pp. 484--489, 2016.

\bibitem{sahoo2019cyber}
S.~Sahoo, T.~Dragi{\v{c}}evi{\'c}, and F.~Blaabjerg, ``Cyber security in control of grid-tied power electronic converters--challenges and vulnerabilities,'' \emph{IEEE Trans. Emerg. Sel. Topics Power Electron.}, vol.~9, pp. 5326--5340, Oct. 2021.

\bibitem{roberts2020}
C.~Roberts, S.-T. Ngo, A.~Milesi, S.~Peisert, D.~Arnold, S.~Saha, A.~Scaglione, N.~Johnson, A.~Kocheturov, and D.~Fradkin, ``Deep reinforcement learning for der cyber-attack mitigation,'' in \emph{2020 IEEE International Conference on Communications, Control, and Computing Technologies for Smart Grids (SmartGridComm)}.\hskip 1em plus 0.5em minus 0.4em\relax IEEE, 2020, pp. 1--7.

\bibitem{WHK}
W.~Kersting, ``Radial distribution test feeders,'' in \emph{IEEE Power Engineering Society Winter Meeting}, 2001.

\bibitem{Cisco_guide}
``Distribution automation feeder automation design guide,'' \url{https://www.cisco.com/c/en/us/td/docs/solutions/Verticals/Distributed-Automation/Feeder-Automation/DG/DA-FA-DG/DA-FA-DG.html}, accessed: [06/07/2023].

\bibitem{woodard2007categorizing}
L.~Woodard, C.~K. Veitch, S.~R. Thomas, and D.~P. Duggan, ``Categorizing threat: building and using a generic threat matrix.'' Sandia National Laboratories (SNL), Albuquerque, NM, and Livermore, CA~…, Tech. Rep., 2007.

\bibitem{Liu2020}
\BIBentryALTinterwordspacing
A.~Liu, J.~Chen, M.~Yu, Y.~Zhai, X.~Zhou, and J.~Liu, ``Watch the unobserved: A simple approach to parallelizing monte carlo tree search,'' in \emph{International Conference on Learning Representations}, 2020. [Online]. Available: \url{https://openreview.net/forum?id=BJlQtJSKDB}
\BIBentrySTDinterwordspacing

\bibitem{bhosekar2018}
A.~Bhosekar and M.~Ierapetritou, ``Advances in surrogate based modeling, feasibility analysis, and optimization: A review,'' \emph{Computers \& Chemical Engineering}, vol. 108, pp. 250--267, 2018.

\bibitem{danihelka2022policy}
\BIBentryALTinterwordspacing
I.~Danihelka, A.~Guez, J.~Schrittwieser, and D.~Silver, ``Policy improvement by planning with gumbel,'' in \emph{International Conference on Learning Representations}, 2022. [Online]. Available: \url{https://openreview.net/forum?id=bERaNdoegnO}
\BIBentrySTDinterwordspacing

\bibitem{deepmind2020jax}
\BIBentryALTinterwordspacing
DeepMind, I.~Babuschkin, K.~Baumli, A.~Bell, S.~Bhupatiraju, J.~Bruce, P.~Buchlovsky, D.~Budden, T.~Cai, A.~Clark, I.~Danihelka, A.~Dedieu, C.~Fantacci, J.~Godwin, C.~Jones, R.~Hemsley, T.~Hennigan, M.~Hessel, S.~Hou, S.~Kapturowski, T.~Keck, I.~Kemaev, M.~King, M.~Kunesch, L.~Martens, H.~Merzic, V.~Mikulik, T.~Norman, G.~Papamakarios, J.~Quan, R.~Ring, F.~Ruiz, A.~Sanchez, L.~Sartran, R.~Schneider, E.~Sezener, S.~Spencer, S.~Srinivasan, M.~Stanojevi\'{c}, W.~Stokowiec, L.~Wang, G.~Zhou, and F.~Viola, ``The {D}eep{M}ind {JAX} {E}cosystem,'' 2020. [Online]. Available: \url{http://github.com/deepmind}
\BIBentrySTDinterwordspacing

\end{thebibliography}


\end{document}